\documentclass[journal]{IEEEtran}
\usepackage{graphicx} 
\usepackage{amsmath,amssymb}
\usepackage{bm}
\usepackage{cancel}

\usepackage{adjustbox}

\usepackage{graphicx}
\usepackage{xcolor}
\usepackage{comment}

\usepackage{tikz}
\usetikzlibrary{plotmarks}
\usepackage{pgfplots}
\usetikzlibrary{calc}
\usetikzlibrary{shapes,arrows}
\usetikzlibrary{decorations.markings}
\usetikzlibrary{positioning}
\pgfplotsset{compat=1.10}
\usetikzlibrary{calc}
\usetikzlibrary{shapes,arrows}
\usetikzlibrary{decorations.markings}
\usetikzlibrary{arrows.meta, positioning}

\newcommand{\Enc}{\mathsf{Enc}}
\newcommand{\Dec}{\mathsf{Dec}}
\newcommand{\Est}{\mathsf{Est}}

\title{Secure Integrated Sensing and Communication: Information Theory Offers Insights
}

\author{Truman Welling,
        Onur G\"unl\"u,
        and Aylin Yener%
        
\thanks{T. Welling is with the Department of Electrical and Computer Engineering, The Ohio State University, Columbus, OH 43210. (e-mail: welling.78@osu.edu).}%
\thanks{O. G\"unl\"u is with the Lehrstuhl f\"ur Nachrichtentechnik, Technische Universit\"at Dortmund, Germany, and also with the Information Theory and Security Laboratory (ITSL), Link\"oping University, Sweden (e-mail: onur.guenlue@tu-dortmund.de).}%
    \thanks{A. Yener is with the Departments of Electrical and Computer Engineering, Computer Science and Engineering, and Integrated Systems Engineering, The Ohio State University, Columbus, OH 43210. (e-mail: yener@ece.osu.edu).}
\thanks{This work was partially supported by the ZENITH Research and Leadership Career Development Fund under Grant ID23.01, EU COST Action 6G-PHYSEC, Swedish Foundation for Strategic Research (SSF) under Grant ID24-0087, and German Federal Ministry of Research, Technology and Space (BMFTR) 6GEM+ Transfer Hub under Grants 16KIS2412 and 16KISS005.}%
}

\begin{document}
\maketitle

{
\begin{abstract}
Integrated sensing and communication (ISAC) combines sensing and communication within a shared system framework by using the same transmitted signal for both objectives. ISAC can improve the efficiency of spectrum and hardware use but also gives rise to new security challenges, as users associated with one function may need to be prevented from inferring information related to the other. This paper surveys information-theoretic approaches to secure ISAC with emphasis on formulations, performance metrics, and fundamental limits. We first review the information-theoretic ISAC models that underlie secure formulations. We then organize the secure ISAC literature according to the protected functionality and the adversary model, covering secure communication, sensing security, and active-adversary settings such as jamming. We also discuss formulations in which communication security and sensing security interact more directly, as well as their connections to privacy and covert communication. Throughout, we highlight the main modeling assumptions and the insights they provide on the tradeoffs among communication reliability, sensing performance, and security.
\end{abstract}
}

\section{Introduction}
Integrated sensing and communication (ISAC), also known as joint communication and sensing, has emerged as a viable candidate for next-generation mobile communication systems. 
ISAC aims to unify the two key operations of future networks that utilize spectrum, namely sensing and communications. As new use cases emerge that utilize both, such as autonomous transportation networks, efficient resource management calls for designs beyond partitioning frequency or co-existence of radar and communications, i.e., true integration that requires a joint waveform and transceiver design architecture. The key to ISAC's success is the ability of the network to automatically react to changing environments thanks to the tight integration of communication and sensing. For instance, a millimeter wave (mmWave) joint communication and radar system can be used to detect a target or to estimate crucial parameters relevant to communication and adapt the communication scheme accordingly~\cite{JCASwithSecurityTutorial}. 

Incorporating communication into a sensing system by embedding communication information into the probing waveform for target sensing introduces new security challenges beyond those in traditional secure communication. In particular, information leakage that occurs between the two functionalities could jeopardize the integrity of either or both operations. For example, a target illuminated for ranging has the ability to gather potentially sensitive information about the transmitted message~\cite{SecureJCASWireless}. As both sensing and secrecy performance are measured with respect to the signal received at the sensed target, there exists a trade-off between the two~\cite{OurJSAITSecureISAC}. 

\begin{figure}[t]
\centering
\includegraphics[keepaspectratio,width=0.6\linewidth]{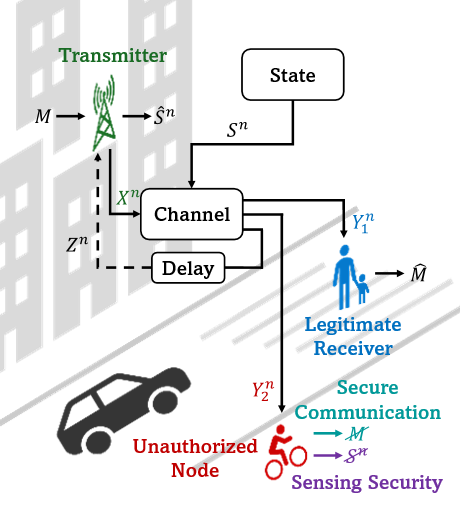}
\vspace{-0.2cm}
\caption{Unified secure ISAC notation used throughout the paper. To communicate a message $M$, a transmitter sends channel inputs $X^n$ to a legitimate receiver, which observes $Y_1^n$, while an unauthorized node observes $Y_2^n$. The sensing-related channel quantity of interest is denoted by $S^n$, and in mono-static ISAC the transmitter may obtain sensing-related feedback $Z^n$ through reflections, since the sensing receiver is co-located with the transmitter. The goal of secure communication is to prevent the unauthorized node from recovering the message $M$, while the goal of secure sensing is to obfuscate the sequence $S^n$. This abstraction and mono-static example capture the common ingredients of the information-theoretic secure ISAC models discussed in the sequel: communication, sensing, feedback, and leakage.}    \label{fig:isac_illustration}
\vspace{-0.3cm}
\end{figure}

{Fig.~\ref{fig:isac_illustration} shows a high-level abstraction of an example secure ISAC system that captures the main elements used throughout this paper, where capital letters denote random variables. In this model, $X^n$ denotes the transmitted signal, $Y_1^n$ and $Y_2^n$ denote the observations at the legitimate receiver and unauthorized node, respectively, $S^n$ represents the sensing-related channel state, and $Z^n$ denotes the sensing feedback obtained as channel-output feedback and available at the sensing receiver. Moreover, we denote the channel state estimate as $\hat{S}^n$ and message estimate as $\hat{M}$. These variables form the basis of the information-theoretic formulations discussed in the sequel.

\begin{figure*}[t]
    \centering
    \includegraphics[width=0.7\linewidth]{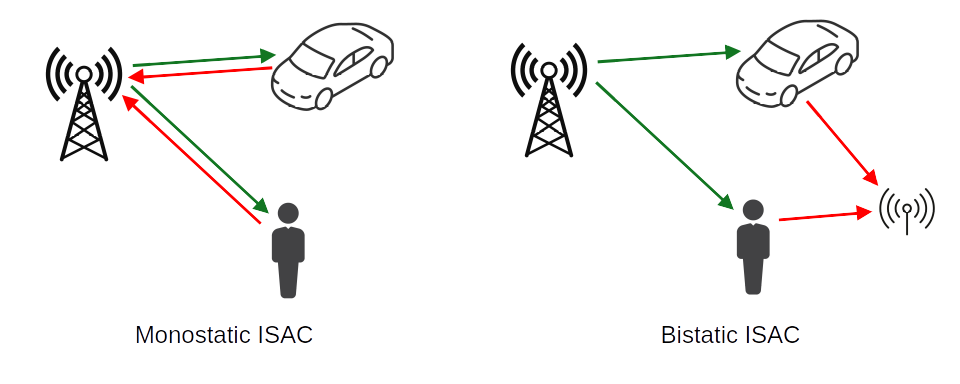}
    \vspace{-0.5cm}
    \caption{In the monostatic ISAC model the transmitter observes the reflected signal, while a separate receiver observes it in the bistatic case. The green arrows depict the transmitted signals and the red arrows the reflections, where the latter are used to perform sensing.}
    \label{fig:monostatic_bistatic_isac_illustration}
    \vspace{-0.4cm}
\end{figure*}

ISAC is founded in physical-layer, which means its fundamental limits are to be determined by information theory. Thus, it is natural to study security from an information-theoretic perspective. While physical-layer security has been extensively investigated for communication systems~\cite{yener2015PLS}, ISAC introduces an additional sensing functionality that is coupled to the same transmitted waveform. This dual use of the signal creates new leakage mechanisms and tradeoffs that require models that jointly capture communication reliability, sensing performance, and security~\cite{OurJSAITSecureISAC}.

From a broader signal-processing and communications perspective, recent surveys have documented the evolution of ISAC as well as the importance of secure and privacy-aware designs; see, e.g.,~\cite{masouros2026ISACEvolution,martins2025delving,su2025integrating,li2025ris}. The focus of this article is complementary to these existing surveys: we emphasize information-theoretic formulations, performance metrics, and fundamental limits, and ultimately draw design insights founded on this information theoretic view. In particular, we emphasize models in which sensing and communication are coupled through common channel inputs and state-dependent observations, rather than treating sensing as a separate or higher-layer functionality. Our aim is not to review all secure ISAC architectures, but rather to convey what is known when security and sensing are jointly modeled at the level of channels, states, rates, and distortions.

Within the context of secure ISAC, secure communication requires that an unauthorized node cannot reliably decode the transmitted message. Sensing security, on the other hand, requires that an unauthorized node cannot accurately estimate or infer the underlying state from its observations. We further distinguish between two broad classes of adversaries: passive adversaries, which only observe the transmission in an attempt to extract sensitive information, and active adversaries, which can additionally interfere with the communication or sensing process.

A central message that emerges from the information-theoretic literature is that sensing can help improve secure communication performance. In particular, sensing-enabled feedback or environmental inference can provide the transmitter with information about propagation conditions and, in some models, support secret-key generation or more informed signaling strategies~\cite{OurJSAITSecureISAC,welling2024transmitterActions,li2025BeamPointing,alhabob2025Predictive}. Moreover, the same coupling that enables such gains also creates new vulnerabilities, since an adversarial node may use its observations not only to decode a message but also to infer sensing-related information~\cite{OurJSAITSecureISAC,guo2026secure}.

Sensing security introduces an additional layer of complexity as it is inherently an estimation problem. Accordingly, secure ISAC models must account for both communication-oriented security metrics and sensing-oriented distortion or inference metrics. This distinction is reflected in formulations that constrain unauthorized reconstruction quality~\cite{chen2025distributionPreservingISAC,masouros2026SecureSensing} as well as in formulations that jointly limit information leakage about communicated messages and sensed information~\cite{ahmadipour2023SecureMessageAndState}. The purpose of this tutorial is to present these formulations, highlight the assumptions that drive the available results, and summarize the main insights that can be drawn from the current information-theoretic literature.}

In this work, two broad classifications of ISAC models are considered, monostatic and bistatic, which indicate the physical location of the sensing receive antenna in relation to the transmit antenna. The two scenarios are depicted in Fig.~ \ref{fig:monostatic_bistatic_isac_illustration}. In a bistatic ISAC scenario, the sensing receiver is physically separated from the transmitter, while in a monostatic ISAC model, the sensing receive antenna is co-located with the transmit antenna.

{
The remainder of this article is organized as follows. Section~\ref{sec:fundamentalISAC} introduces the common notation, performance metrics, and high-level model categories used throughout the paper. In Section~\ref{sec:attackmodelsforsecureISAC}, we present the information-theoretic secure ISAC models, including both communication and sensing security. Section~\ref{sec:NewDirections} discusses formulations in which communication security and sensing security interact more directly, together with their connections to privacy and covert communication. Section~\ref{sec: Discussions} concludes with future directions. With this organization, we intend to separate the common modeling framework from the individual papers, so that each paper can be understood as a particular secure ISAC variant.
}

\begin{table*}[t]
\caption{Taxonomy of the secure ISAC formulations discussed in this paper.}
\label{tab:secure_isac_taxonomy}
\centering
\renewcommand{\arraystretch}{1.1}
\begin{tabular}{|p{2.8cm}|p{2.2cm}|p{2.2cm}|p{2.4cm}|p{3.2cm}|}
\hline
Category & Representative works & Protected quantity & Performance metrics & Modeling Assumptions\\
\hline
\hline
Secure communication &
\cite{OurJSAITSecureISAC,welling2024transmitterActions,li2025BeamPointing,gunlu2023secureBinaryAWGN,gunlu2024nonasymptotic,gong2024UntrustedSensing}
& Message &
Secure rate, leakage, distortion &
Feedback, state dependence, transmitter actions, trusted/untrusted sensing node \\
\hline
Sensing security &
\cite{chen2025distributionPreservingISAC,masouros2026SecureSensing}
& State / target information &
Reconstruction distortion, sensing ambiguity &
State knowledge, shared randomness, waveform design \\
\hline
Active adversaries &
\cite{xu2024jammingIRS,liu2024gameJamming}
& Reliability and robustness under attack &
SNR, utility, robustness &
Jamming, RIS control\\
\hline
Joint message and sensing security &
\cite{ahmadipour2023SecureMessageAndState,guo2026secure}
& Message and state &
Mutual-information leakage, rate-distortion tradeoff &
Noncausal state knowledge vs.\ feedback-based sensing \\
\hline
Covert ISAC &
\cite{BlochCovertJCAS}
& Undetectability of communication &
Detection constraint, sensing error exponent &
Warden observation, reflected channel, fixed parameter $\theta$ \\
\hline
\end{tabular}
\vspace{-0.3cm}
\end{table*}

\section{Information-theoretic Models for ISAC}\label{sec:fundamentalISAC}
{Information-theoretic studies of ISAC build on channel models in which the transmitted waveform simultaneously supports reliable communication and state sensing~\cite{Zhang_2011,AcademicsJCASTutorial,MassiveMIMOforJCAS}. Throughout this paper, we use the following basic terminology. The transmitter sends channel inputs to a legitimate receiver while also obtaining sensing-related information either through reflected signals in monostatic ISAC or through a separate sensing receiver in bistatic ISAC. The random variable $S$ denotes the environmental state or sensing-related quantity of interest, such as propagation conditions or target-dependent parameters, and its precise interpretation depends on the model under consideration.

The main performance metrics we consider are communication rate, secure communication rate, sensing distortion, and information leakage. In secure communication models discussed in the sequel, secrecy is expressed either through a mutual-information leakage constraint or through a variational-distance constraint. In the sensing-security models, the adversary's performance is subject to a distortion or inference constraint.

A foundational information-theoretic ISAC model is developed in~\cite{MariCaireKramerISACfirst}, where messages are encoded and sent over a state-dependent channel with generalized feedback so that the transmitter can both communicate reliably and estimate the channel state from the transmitted codewords and the observed feedback. For memoryless channels with independent and identically distributed (i.i.d.) states,~\cite{MariCaireKramerISACfirst} characterizes the optimal tradeoff between communication rate and state-estimation distortion. This formulation has since been extended to multiple-access channels~\cite{MariMACJCAS} and broadcast channels~\cite{GiuseppeITJSACJournal}, and it provides the information-theoretic baseline from which the secure ISAC models discussed next are derived.

The secure ISAC literature considered in this article is organized according to the protected functionality and the adversary model: Section~\ref{sec:attackmodelsforsecureISAC} treats passive and active adversaries, while Section~\ref{sec:NewDirections} discusses formulations in which communication security, sensing security, privacy, and covertness interact.

The models reviewed in this work differ mainly in (i) what must be protected, (ii) what the transmitter knows about the channel, (iii) how sensing information is obtained, and (iv) whether the adversary is passive (observation-only) or active (able to interfere). The protected object can be the communicated message, the sensing-related state, or both. State knowledge can be noncausal, causal through feedback, or partly controllable through transmitter actions. Sensing can be monostatic, where the transmitter learns from reflections, or bistatic, where sensing is carried out by a separate node. Finally, the adversary can be passive, in which case security is expressed through leakage or inference constraints, or active, in which case robustness against jamming or strategic interference becomes part of the formulation. Table~\ref{tab:secure_isac_taxonomy} summarizes the main categories of secure ISAC models discussed in this paper, the protected quantities, the primary performance metrics, and the key modeling assumptions that distinguish them.

}

\section{Fundamental Limits of Secure ISAC}\label{sec:attackmodelsforsecureISAC}
{
In an ISAC system, the communicated message and/or the sensing-related information may be sensitive. Communication security aims to limit what an unauthorized node can learn about the transmitted message, whereas sensing security aims to limit what an unauthorized node can infer about the environmental state, target, or sensing outcome. In this section, we review the information-theoretic secure-ISAC models in the literature for passive and active adversaries.
}

\subsection{Passive Adversaries}
A passive adversary gathers information about the overheard signals without attempting to influence the exchange of information. For ISAC systems, this includes both communication and sensing information. The inherent differences in communication and sensing, and in the respective approaches to secure them, render it useful to consider them separately.


\subsubsection{Secure Communication}
{
We first consider passive-adversary models in which the communicated message must be protected from an unauthorized node. In the general exposition, we use the term unauthorized node for a generic attacker. In model-specific discussions, we retain terms such as eavesdropper when they are part of the original formulation.

\begin{figure*}[t]
  \centering
  \resizebox{.7\linewidth}{!}{
    \begin{tikzpicture}
      \node (a) at (-.5,-1.6) [draw,rounded corners = 6pt, minimum width=2.2cm,minimum height=0.8cm, align=left] {$\widehat{S}^n_j = \Est_j(X^n,Z^n),~{j\in\{1,2\}}$};
      \node (c) at (4.5,-3.1) [draw,rounded corners = 5pt, minimum width=1.3cm,minimum height=0.6cm, align=left] {$P_{Y_1Y_2Z|S_1S_2X}$};
      \node (state) at (7.5,-3.1) [draw,rounded corners = 5pt, minimum width=1.3cm,minimum height=0.6cm, align=left] {$P_{S_1S_2}$};
      \draw[decoration={markings,mark=at position 1 with {\arrow[scale=1.5]{latex}}},
      postaction={decorate}, thick, shorten >=1.4pt] ($(state.west)+(0,0.15)$) -- ($(c.east)+(0,0.15)$);
      \draw[decoration={markings,mark=at position 1 with {\arrow[scale=1.5]{latex}}},
      postaction={decorate}, thick, shorten >=1.4pt] ($(state.west)+(0,-0.15)$) -- ($(c.east)+(0,-0.15)$);
      \node (b) at (5.5,-1) [draw,rounded corners = 6pt, minimum width=2.2cm,minimum height=0.8cm, align=left] {$\widehat{M}=\Dec(Y_1^n,S_1^n)$};
      \node (g) at (4.5,-4.9) [draw,rounded corners = 5pt, minimum width=1cm,minimum height=0.6cm, align=left] {Eve};
      \node (z) at (-.5,-2.9) [draw, dashed, minimum width =6.25cm, minimum height=4.4cm, align=center] {};
      \node (z1) at (-.5,-.4) {Transmitter};
      \draw[decoration={markings,mark=at position 1 with {\arrow[scale=1.5]{latex}}},
      postaction={decorate}, thick, shorten >=1.4pt] ($(state.west)+(-0.2,0.15)$) -- ($(b.south)+(1.1,0)$);
      \node (s1) at (6.1,-2.7) {$S_{1,i}$};
      \node (s2) at (6.1,-3.56) {$S_{2,i}$};
      \draw[decoration={markings,mark=at position 1 with {\arrow[scale=1.5]{latex}}},
      postaction={decorate}, thick, shorten >=1.4pt] ($(state.west)+(-0.2,-0.15)$) -- ($(b.south)+(1.15,-3.5)$) -- ($(g.east)+(0,0.00)$);
      \node (a1) [below of = a, node distance = 1.5cm] {$X_i$};
      \draw[decoration={markings,mark=at position 1 with {\arrow[scale=1.5]{latex}}},
      postaction={decorate}, thick, shorten >=1.4pt] ($(c.north)+(0.0,0)$) -- ($(b.south)-(1,0)$) node [midway, right] {$Y_{1,i}$};
      \draw[decoration={markings,mark=at position 1 with {\arrow[scale=1.5]{latex}}},
      postaction={decorate}, thick, shorten >=1.4pt] (a1.east) -- ($(c.west)-(0,0.0)$);
      \draw[decoration={markings,mark=at position 1 with {\arrow[scale=1.5]{latex}}},
      postaction={decorate}, thick, shorten >=1.4pt,dashed] (a1.north) -- ($(a.south)-(0,0.0)$);
      \draw[decoration={markings,mark=at position 1 with {\arrow[scale=1.5]{latex}}},
      postaction={decorate}, thick, shorten >=1.4pt] ($(c.north)-(0.5,0.0)$) -- ($(c.north)-(0.5,-0.3)$) -- ($(c.north)-(6,-0.3)$) -- ($(a.south)-(1,0)$);
      \draw[decoration={markings,mark=at position 1 with {\arrow[scale=1.5]{latex}}},
      postaction={decorate}, thick, shorten >=1.4pt] (c.south) -- (g.north) node [midway, right] {$Y_{2,i}$};
      \node (b2) [right of = b, node distance = 4cm] {$\widehat{M}=\big(\widehat{M}_1,\widehat{M}_2\big)$};
      \draw[decoration={markings,mark=at position 1 with {\arrow[scale=1.5]{latex}}},
      postaction={decorate}, thick, shorten >=1.4pt] (b.east) -- (b2.west);
      \node (a2) [below of = a, node distance = 4.4cm] {$M=(M_1,M_2)$};
      \node (f2) at (-.5,-4.3) [draw,rounded corners = 5pt, minimum width=1cm,minimum height=0.6cm, align=left] {$X_i=\Enc_i(M,Z^{i-1})$};
      \draw[decoration={markings,mark=at position 1 with {\arrow[scale=1.5]{latex}}},
      postaction={decorate}, thick, shorten >=1.4pt]  (f2.north) -- (a1.south);
      \draw[decoration={markings,mark=at position 1 with {\arrow[scale=1.5]{latex}}},
      postaction={decorate}, thick, shorten >=1.4pt] (a2.north) -- (f2.south) ;
      \draw[decoration={markings,mark=at position 1 with {\arrow[scale=1.5]{latex}}},
      postaction={decorate}, thick, shorten >=1.4pt] ($(c.north)-(6,-0.3)$) -- ($(f2.north)-(1,0)$) node [pos=0.0, left] {$Z_{i-1}$};
    \end{tikzpicture}
  }
  \vspace{-0.2cm}
  \caption{ISAC model under partial secrecy from \cite{OurJSAITSecureISAC}, where only $M_2$ should be kept secret from Eve. {Here, $j\in\{1,2\}$ indexes the two channel state components $S_1$ and $S_2$, and $Z^{i-1}$ denotes the channel-output feedback available at the transmitter up to time $i-1$ (i.e., with a unit time delay as compared to the transmitted index $X_i$) for $i=1,2,\ldots, n$.}}
\label{fig:SecureJCASModel}
\vspace{-0.3cm}
\end{figure*}
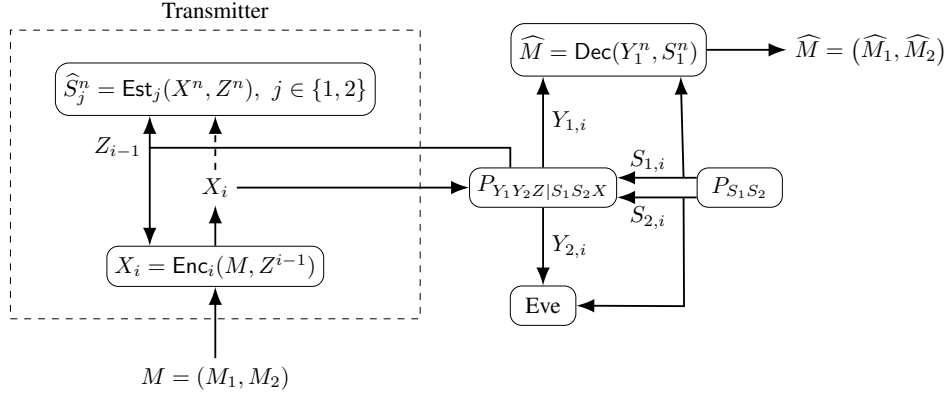

A canonical information-theoretic formulation is given in~\cite{OurJSAITSecureISAC}. In this model, a transmitter communicates with a legitimate receiver over a state-dependent broadcast channel in the presence of an eavesdropper, while simultaneously estimating sensing-related states from channel-output feedback. The message is split into a public part $M_1$ and a confidential part $M_2$, and the transmitter aims to communicate reliably to the legitimate receiver, keep $M_2$ secret from the eavesdropper, and estimate the relevant state sequences subject to distortion constraints. The proposed secure ISAC model can be viewed as an extension of the wiretap channel with feedback models, e.g., ~\cite{AhlswedeCaiWTCwithFeedback,AsafCohenWTCwithFeedback}.

The model in~\cite{OurJSAITSecureISAC} is depicted in Fig.~\ref{fig:SecureJCASModel}. The state is written as $S=(S_1,S_2)$, where $S_1$ and $S_2$ denote the state components associated with the legitimate-receiver and eavesdropper links, respectively. The legitimate receiver observes $(Y_1^n,S_1^n)$, the eavesdropper observes $(Y_2^n,S_2^n)$, and the transmitter observes delayed channel-output feedback $Z^{i-1}$ causally for $i=1,2,\ldots, n$. The legitimate receiver produces an estimate $\hat{M}$ of the message based on its observation. The auxiliary random variables $U$ and $V$ describe the coding layers used in the achievability scheme: $U$ is associated with the public layer and $V$ with the confidential layer.

A tuple $(R_1,R_2,D_1,D_2)$ is achievable if there exists a sequence of blocklength-$n$ codes such that decoding error probability at the legitimate receiver tends to zero, confidential-message leakage satisfies
\begin{align}
    \lim_{n\rightarrow\infty} I(M_2;Y_2^n,S_2^n)= 0
\end{align}
and expected distortions satisfy, for $j\!\in\!\{1,2\}$,
\begin{align}
\limsup_{n\to\infty}\mathbb{E}\big[d_j(S_j^n,\widehat{S}_j^n)\big]\leq D_j.
\end{align}
Thus, strong secrecy here means that the total information leakage about the confidential message vanishes asymptotically, while the distortion constraints ensure that the state estimates satisfy the required sensing performance. According to these definitions, for the perfect channel-output feedback case,~\cite{OurJSAITSecureISAC} showed that the achievable strong secrecy--distortion region includes tuples $(R_1,R_2,D_1,D_2)$ satisfying
\begin{align}
	R_1 &\leq I(U;Y_1|S_1),\label{eq:secure_comm_common_rate}\\
	R_2 &\leq \min\{R_2',I(V;Y_1|S_1)\},\label{eq:secure_comm_secure_rate}\\
    D_j &\geq\mathbb{E}\big[d_j(S_j,\widehat{S}_j)\big] \qquad \textnormal{for } j=1,2, \label{eq:gunlu_distortion_constraints}
\end{align}
where $R_1$ and $R_2$ are the public and confidential message rates, respectively, and
\begin{align}
    R_2'&= \big[I(V;Y_1|S_1,U)-I(V;Y_2|S_2,U)\big]^+\nonumber\\
    &\qquad\quad+ H(Y_1|Y_2,S_2,V), \label{eq:gunlu_secure_rate}
\end{align}
which contains two distinct contributions to the confidential-message rate. The first term in \eqref{eq:gunlu_secure_rate} corresponds to an expanded wiretap-coding contribution and relies on the legitimate receiver having a stronger observation than the eavesdropper~\cite{OurJSAITSecureISAC}. In contrast, the second term, $H(Y_1|Y_2,S_2,V)$ represents secret-key material extracted from sensing feedback, which does not require such an advantage and is used in a block-Markov coding scheme to protect subsequent transmissions. This decomposition makes explicit how sensing-related feedback can create additional secrecy resources beyond classical channel-advantage mechanisms.

The same work fully characterizes the secrecy-distortion regions for physically degraded and reversely physically degraded ISAC channels. In the physically degraded case, the eavesdropper's observation is a stochastically degraded version of the legitimate receiver's observation, so the legitimate receiver is the stronger terminal from the standpoint of observation quality. In the reversely physically degraded case, the opposite ordering holds. In the former case, the auxiliary random variable $U$ is not needed; in the latter case, the wiretap-coding contribution becomes zero and secrecy is obtained entirely through the secret-key term. These two degraded cases therefore isolate the two distinct mechanisms by which secure communication is supported in the model.

These results show that sensing-related feedback can improve secure communication rates and can even enable secure communication in settings where wiretap coding alone would not suffice; see also \cite{Yingyaosecurefeedback} for the first neural code designs for such feedbacked wiretap coding settings. The model relies on strong assumptions, including a memoryless channel law and idealized feedback, so it should be viewed as a baseline information-theoretic formulation rather than a complete engineering model.

\paragraph*{Transmitter actions}
The model in~\cite{welling2024transmitterActions} allows transmitter actions that affect the channel-state distribution. This captures settings in which the transmitter can influence the sensed environment, for example through controlled motion or configuration changes. In this case, the state is no longer independent of the transmitted sequence, and the achievable rate expressions change accordingly. In particular, the equality $I(V;S_1)=0$, used in the derivation of~\eqref{eq:secure_comm_common_rate}-\eqref{eq:gunlu_secure_rate} for the state-independent model, no longer holds. As a result, the state enters the mutual-information terms directly rather than only through conditioning.

\paragraph*{Binary-input AWGN model}
The work in~\cite{gunlu2023secureBinaryAWGN} studies a binary-input additive white Gaussian noise secure-ISAC channel. For channel input $X_i\in\{-\sqrt{P},\sqrt{P}\}$ at time $i$ with transmit power $P$, the outputs are
\begin{align}
    Y_{1,i} = S_{1,i}X_i + N_{1,i},\\
    Y_{2,i} = S_{2,i}X_i + N_{2,i},
\end{align}
where $S_1,S_2\in\{-a,a\}$ with amplitude $a\geq 0$ are i.i.d. according to $P_{S_1,S_2}$ and $N_{1,i}$ and $N_{2,i}$ are zero-mean Gaussian noise with variances $\sigma_1^2$ and $\sigma_2^2$, respectively. Under full secrecy and Hamming distortion metrics,~\cite{gunlu2023secureBinaryAWGN} derives an outer bound on the secrecy-distortion region:
\begin{align}
    R & \leq\min\{R',\mathbb{E}_{S_1}[h(S_1X+N_1)]\}\nonumber\\
    &\qquad\qquad-\frac{1}{2}\log(2\pi e\sigma_1^2), \label{eq:BAWGNrate}\\
    D_j& \geq Q\bigg(\frac{a\sqrt{P}}{\sigma_j}\bigg)\qquad \text{for } j=1,2
\end{align}
with
\begin{align}
    R'&= \mathbb{E}_{S_1S_2}[h(S_1X+N_1|S_2X+N_2)]\nonumber\\
    & \qquad\quad+ \mathbb{E}_{X}[h(S_1X+N_1|S_2)].\label{eq:BAWGNsecureRate}
\end{align}
This outer bound is tight for degraded secure ISAC channels. Unlike~\cite{OurJSAITSecureISAC,welling2024transmitterActions}, the secrecy requirement here is formulated using weak secrecy, meaning that the leakage normalized by the blocklength $n$ vanishes asymptotically.

\paragraph*{Finite-blocklength regime}
Low-latency secure ISAC is considered in~\cite{gunlu2024nonasymptotic}, which studies a nonasymptotic version of the models considered before. The secrecy condition is expressed through a variational-distance constraint, which provides strong secrecy and leads to nonasymptotic inner bounds, i.e., the blocklength $n$ is finite.

\paragraph*{Additional recent directions}
Two further directions are represented by~\cite{li2025BeamPointing}, which studies binary beampointing secure ISAC with block memory and feedback, and~\cite{gong2024UntrustedSensing}, which studies a bistatic setting with an untrusted sensing node. These papers broaden the secure-communication picture beyond the original memoryless monostatic formulation and show that the unauthorized observer can also be the sensing terminal itself.

Thus, the main modeling choices in secure communication for ISAC are the source of feedback, the role of the state, whether the transmitter can influence that state, and which node is considered unauthorized. Across these models, feedback and state knowledge emerge as the main enablers of secure communication, while changes in the state model determine whether secrecy comes primarily from wiretap coding or from key extraction. This highlights that sensing is not merely an auxiliary functionality, but can actively contribute to enabling secure communication.
}

\subsubsection{Secure Sensing}
{
We next consider the case in which the protected object is not the communicated message, but the sensing-related information itself. In this setting, an unauthorized node uses its observations to infer the environmental state, target characteristics, or target location. Accordingly, sensing security is formulated by limiting the quality of this unauthorized inference rather than by requiring message secrecy. A representative sensing-security setting is illustrated in Fig.~\ref{fig:sensing_security_model}.

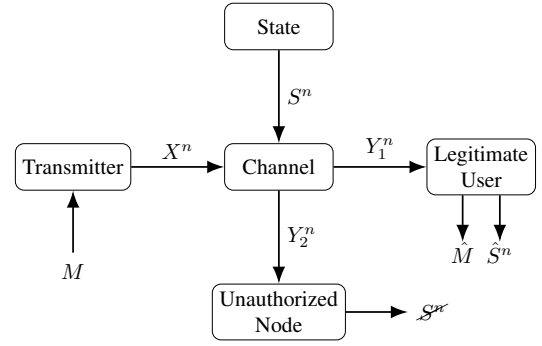
\begin{figure}[t]
\centering
\resizebox{0.8\columnwidth}{!}{
\begin{tikzpicture}[
    node distance=1.5cm,
    every node/.style={align=center},
    latexarrow/.style={
        thick,
        decoration={
            markings,
            mark=at position 1 with {\arrow[scale=1.5]{latex}}
        },
        postaction={decorate},
        shorten >=1.4pt
    },
    slashlatexarrow/.style={
        thick,
        decoration={
            markings,
            mark=at position 1 with {\arrow[scale=1.5]{latex}}
        },
        postaction={decorate},
        shorten >=1.4pt
    }
]

\node[draw, rectangle, rounded corners, minimum width=1.75cm, minimum height=0.75cm]
    (encoder) {Transmitter};
\node[draw=none, inner sep=5pt] (m) [below=1.0cm of encoder] {$M$};
\node[draw, rectangle, rounded corners, minimum width=1.75cm, minimum height=0.75cm]
    (channel) [right=of encoder] {Channel};
\node[draw, rectangle, rounded corners, minimum width=1.75cm, minimum height=0.75cm]
    (state) [above=of channel] {State};
\node[draw, rectangle, rounded corners, minimum width=1.75cm, minimum height=0.75cm]
    (decoder) [right=of channel] {Legitimate\\ User};
\node[draw, rectangle, rounded corners, minimum width=1.75cm, minimum height=0.75cm]
    (unauth) [below=of channel] {Unauthorized\\ Node};
\node[draw=none, inner sep=5pt]
    (s) [right=1.0cm of unauth] {$\cancel{S^n}$};

\node[draw=none, inner sep=0pt]
    (mhat) [below=0.75cm of decoder, xshift=-0.3cm] {$\hat{M}$};
\node[draw=none, inner sep=0pt]
    (shat) [below=0.75cm of decoder, xshift=0.3cm] {$\hat{S}^n$};

\draw[latexarrow] (m) -- (encoder.south);
\draw[latexarrow] (encoder) -- node[above, draw=none] {$X^n$} (channel);
\draw[latexarrow] (channel) -- node[above, draw=none] {$Y_1^n$} (decoder);
\draw[latexarrow] (channel) -- node[right, draw=none] {$Y_2^n$} (unauth);
\draw[latexarrow] (state) -- node[right, draw=none] {$S^n$} (channel);

\draw[latexarrow] (unauth) -- (s);
\draw[latexarrow] ($(decoder.south)+(-0.3,0)$) -- (mhat.north);
\draw[latexarrow] ($(decoder.south)+(0.3,0)$) -- (shat.north);

\end{tikzpicture}
}
\vspace{-0.1cm}
\caption{A representative sensing-security setting, where channel-output (sensing) feedback is not considered for simplicity. The transmitter and legitimate system cooperate to sense the channel state or a target-related quantity, while an unauthorized node attempts to reconstruct this sensing-related information from its own observations. The security objective is to degrade the quality of this unauthorized inference.}
\label{fig:sensing_security_model}
\vspace{-0.3cm}
\end{figure}

A natural baseline is the point-to-point state-sensing formulation of~\cite{MariCaireKramerISACfirst}, where communication and state estimation are already coupled. Building on this viewpoint,~\cite{chen2025distributionPreservingISAC} studies a model in which the transmitter communicates with and senses a legitimate receiver while an unauthorized node attempts to reconstruct the sensed state. The transmitter has noncausal access to a noisy version of the state, the legitimate parties share common randomness, and the unauthorized node can additionally receive rate-limited side information from a henchman. The security requirement is expressed as a lower bound on the distortion between the transmitter's reconstruction $\hat{S}^n$ and the unauthorized reconstruction $\hat{S}_E^n$:
\begin{align}
    \mathbb{E}\big[d(\hat{S}^n,\hat{S}_E^n)\big]\geq D_E.
\end{align}
This formulation shows that sensing security is naturally an estimation-theoretic objective.

A complementary direction is given in~\cite{masouros2026SecureSensing}, which impairs unauthorized sensing through ambiguity-function engineering. The key idea is to introduce controlled imperfections into the transmitted waveform so as to create ghost targets in the unauthorized range profile, while allowing the legitimate system to suppress them using matched filtering. Thus, sensing security can also be pursued without requiring explicit knowledge of the unauthorized sensor.

These works illustrate a basic distinction between communication security and sensing security. In communication security, the aim is to hide a message from an unauthorized decoder. In sensing security, the aim is to degrade an unauthorized estimator or detector. This difference naturally leads to different metrics, coding strategies, and design tradeoffs.
}

\subsection{Active Adversaries}
{
An active adversary does not merely observe the system, but also attempts to degrade communication or sensing performance. In the current secure-ISAC literature, the dominant active-adversary model is jamming. The main design objective is therefore to maintain communication reliability and sensing performance in the presence of deliberate interference, possibly while adapting the signaling strategy to the adversary's behavior.

\paragraph*{Anti-jamming design via joint beamforming, reconfigurable intelligent surface (RIS)} \cite{xu2024jammingIRS} considers communication with multiple users and sensing of a blocked target in the presence of a jammer. The optimization variables are the transmit beamformer, the phase matrix of an aerial RIS, and the deployment location of that RIS. The sensing requirement is enforced through a minimum reflected-signal SNR at the base station. This line of work represents an optimization-based robust-design viewpoint in which communication, sensing, and infrastructure placement are co-designed against active attacks.

\paragraph*{Game-theoretic anti-jamming beamforming}
The work in~\cite{liu2024gameJamming} models the interaction between the transmitter and the jammer as a Stackelberg game. A multi-antenna base station communicates with a legitimate user and senses a target using a common waveform, while the jammer observes the transmitted signal and chooses its jamming power strategically. The base station first selects a precoding matrix subject to power and sensing constraints, and the jammer then best-responds. This formulation is useful because it captures the fact that an active adversary may adapt to the legitimate system rather than act in a fixed, nonresponsive way.

\paragraph*{Predictive secure beamforming against aerial eavesdroppers}
A complementary direction is given in~\cite{alhabob2025Predictive}, which studies secure ISAC with multiple aerial eavesdroppers whose channels are predicted and tracked. One main point of this work relevant for this tutorial is that sensing can also support secure communication against mobile threats by helping predict the adversary's channel evolution and thereby improve beamforming decisions.

Thus, the active-adversary literature shows that, once adversaries are allowed to interfere or move strategically, secure ISAC becomes a joint robustness-and-adaptation problem.
}
      
\section{Synergies in Secure ISAC}\label{sec:NewDirections}

\subsection{Sensing and Communication Synergies}
{
The simultaneous protection of communicated information and sensing-related information has received comparatively limited attention in the information-theoretic literature. A first formulation is given in~\cite{ahmadipour2023SecureMessageAndState}, where the transmitter observes the state sequence noncausally and communicates both a message and state information to a legitimate user while keeping both the transmitted message and a function of the channel state sequence secure from an eavesdropper. The achievability scheme is based on superposition coding, with one layer carrying general state information and another carrying the secure message and the secure function of the state.

A second recent direction is~\cite{guo2026secure}, which considers a monostatic secure-ISAC setting in which the transmitter simultaneously sends a confidential message to a legitimate receiver and senses an environmental state while a passive adversary attempts to recover both the message and the sensed information. This result makes the joint protection problem explicit in a feedback-based monostatic setting and highlights how key extraction via feedback shapes the tradeoff between message secrecy and sensing-related performance.

Overall, these works suggest that secure communication and sensing security should not always be treated as separate add-ons. In some models, the two objectives are structurally coupled, and the form of that coupling depends strongly on whether the state is known in advance or must be inferred through sensing. 
}

\subsection{Sensing Security and ISAC Privacy}
{
The information-theoretic formulations of sensing security discussed in this paper use different metrics. The distortion-based formulation in Section~\ref{sec:attackmodelsforsecureISAC} limits the quality of an unauthorized reconstruction~\cite{chen2025distributionPreservingISAC,masouros2026SecureSensing}, whereas the joint message-and-state formulation in \cite{ahmadipour2023SecureMessageAndState} limits information leakage through a mutual-information criterion. These metrics can each be operationally meaningful.

This observation also clarifies the connection between sensing security and privacy. Preventing an unauthorized node from accurately inferring a user's location or state provides one form of privacy, but ISAC privacy can be broader than that. In particular, privacy may also need to be guaranteed against a legitimate sensing receiver rather than only against an external eavesdropper. One way to model such a requirement is to limit the sensing resolution or reconstruction fidelity that the legitimate system is allowed to achieve with respect to user-sensitive state information~\cite{masouros2026SecureSensing}.

Accordingly, sensing security and ISAC privacy should be viewed as related but distinct notions. Sensing security focuses on restricting unauthorized inference, whereas privacy may additionally require limiting what even authorized sensing entities are allowed to infer.
}

\subsection{Covert ISAC}
{
In covert communication, the objective is not only to convey information reliably, but also to prevent a warden from detecting that communication is taking place. In ISAC, this detection problem coexists with a sensing task, which creates an additional interaction between communication concealment and environment inference.

The information-theoretic covert-ISAC model in~\cite{BlochCovertJCAS} contains three channels: the legitimate receiver's channel, the warden's channel, and a reflection channel observed by the transmitter. All three depend on a parameter $\theta$, representing the fixed channel or environment parameter that the transmitter seeks to estimate from its inputs and reflected observations. The warden is assumed to know $\theta$ and tries to decide whether communication is occurring, while the transmitter aims to communicate covertly and to sense $\theta$.

Unlike the distortion-based sensing formulations discussed earlier,~\cite{BlochCovertJCAS} measures sensing performance through an error exponent, because $\theta$ is fixed during the transmission. A notable conclusion is that, for a binary input alphabet consisting of one benign symbol and one information-carrying symbol, there need not be a tradeoff between covert communication rate and sensing performance. For this tutorial, the main point of this work is that covert ISAC introduces a third type of protection objective: besides reliability and secrecy, the system may also need to hide the presence of communication. Much work remains in this area beyond this model.
}

 \section{Future Directions} \label{sec: Discussions}

{
Secure ISAC remains an emerging area, especially from a fundamental-limits perspective. The broader ISAC literature increasingly recognizes that communication, sensing, security, and privacy must be treated jointly rather than sequentially. This creates a clear role for information-theoretic analysis: it can identify which tradeoffs are intrinsic to the dual functionality of ISAC and which are artifacts of a particular architecture or algorithmic design.
}

A multitude of open directions remain unexplored. Multi-transmitter and multi-receiver systems and their joint design of transmission strategies towards fundamental limits are a clear direction. Capacity results in channel models with feedback and state are available only with strong assumptions that could exclude practical ISAC models, and as such, models with generalized feedback tailored to ISAC as well as  transmitter induced channels (for example through movement) present a new direction. Likewise, directions where machine learning can play a role in learning the environment and in turn designing the signaling (and coding) strategies are those that can aid and/or complement the information theoretic studies.

A relatively unexplored role is that of time-varying channels. Care must be exercised to consider models valid at time scales relevant for ISAC and to investigate whether channel variations can be exploited to protect conveyed or sensed information.

{
Dynamically evolving states are another important direction for secure ISAC. Open-loop designs may exhibit markedly different tradeoffs depending on how sensing and communication resources are allocated. In particular, time-sharing and simultaneous sensing-and-communication strategies can lead to different operating regimes, and understanding how secrecy and privacy constraints modify these regimes remains an open problem for fundamental-limits analysis.

An important direction is to understand when secure ISAC can be achieved without relying on large amounts of pre-shared randomness or idealized coordination assumptions. While artificial randomness can improve security or privacy, it introduces additional implementation complexity. From a fundamental-limits perspective, it is thus important to identify scenarios in which feedback, environmental learning, or channel properties can provide inherent security guarantees.
}


\vspace{-0.3cm}
\bibliographystyle{IEEEtran}
\bibliography{references2026}

\begin{thebibliography}{10}
\providecommand{\url}[1]{#1}
\csname url@samestyle\endcsname
\providecommand{\newblock}{\relax}
\providecommand{\bibinfo}[2]{#2}
\providecommand{\BIBentrySTDinterwordspacing}{\spaceskip=0pt\relax}
\providecommand{\BIBentryALTinterwordstretchfactor}{4}
\providecommand{\BIBentryALTinterwordspacing}{\spaceskip=\fontdimen2\font plus
\BIBentryALTinterwordstretchfactor\fontdimen3\font minus
  \fontdimen4\font\relax}
\providecommand{\BIBforeignlanguage}[2]{{%
\expandafter\ifx\csname l@#1\endcsname\relax
\typeout{** WARNING: IEEEtran.bst: No hyphenation pattern has been}%
\typeout{** loaded for the language `#1'. Using the pattern for}%
\typeout{** the default language instead.}%
\else
\language=\csname l@#1\endcsname
\fi
#2}}
\providecommand{\BIBdecl}{\relax}
\BIBdecl

\bibitem{JCASwithSecurityTutorial}
Z.~Wei, F.~Liu, C.~Masouros, N.~Su, and A.~P. Petropulu, ``Toward
  multi-functional {6G} wireless networks: {Integrating} sensing,
  communication, and security,'' \emph{IEEE Communications Magazine}, vol.~60,
  no.~4, pp. 65--71, 2022.

\bibitem{SecureJCASWireless}
N.~Su, F.~Liu, and C.~Masouros, ``Secure radar-communication systems with
  malicious targets: {I}ntegrating radar, communications and jamming
  functionalities,'' \emph{IEEE Transactions on Wireless Communications},
  vol.~20, no.~1, pp. 83--95, Jan. 2021.

\bibitem{OurJSAITSecureISAC}
O.~Günlü, M.~R. Bloch, R.~F. Schaefer, and A.~Yener, ``Secure integrated
  sensing and communication,'' \emph{IEEE Journal on Selected Areas in
  Information Theory}, vol.~4, pp. 40--53, 2023.

\bibitem{yener2015PLS}
A.~Yener and S.~Ulukus, ``Wireless physical-layer security: Lessons learned
  from information theory,'' \emph{Proc. IEEE}, vol. 103, no.~10, pp.
  1814--1825, 2015.

\bibitem{masouros2026ISACEvolution}
D.~Zhang, Y.~Cui, X.~Cao, N.~Su, Y.~Gong, F.~Liu, W.~Yuan, X.~Jing,
  J.~Andrew~Zhang, J.~Xu, C.~Masouros, D.~Niyato, and M.~Di~Renzo, ``Integrated
  sensing and communications over the years: An evolution perspective,''
  \emph{IEEE Communications Surveys \& Tutorials}, vol.~28, pp. 5014--5048,
  2026.

\bibitem{martins2025delving}
{\'O}.~G. Martins, H.~{\AA}kesson, M.~Gomes, D.~P. Osorio, P.~Sen, and J.~P.
  Vilela, ``Delving into security and privacy of joint communication and
  sensing: A survey,'' \emph{IEEE Open Journal of the Communications Society},
  2025.

\bibitem{su2025integrating}
N.~Su, F.~Liu, J.~Zou, C.~Masouros, G.~C. Alexandropoulos, A.~Mourad, J.~L.
  Hernando, Q.~Zhang, and T.-T. Chan, ``Integrating sensing and communications
  in {6G? Not} until it is secure to do so,'' \emph{arXiv preprint
  arXiv:2503.15243}, 2025.

\bibitem{li2025ris}
Y.~Li, F.~Khan, M.~Ahmed, A.~A. Soofi, W.~U. Khan, C.~K. Sheemar, M.~Asif, and
  Z.~Han, ``{RIS}-based physical layer security for integrated sensing and
  communication: {A} comprehensive survey,'' \emph{IEEE Internet of Things
  Journal}, 2025.

\bibitem{welling2024transmitterActions}
T.~Welling, O.~G{\"u}nl{\"u}, and A.~Yener, ``Transmitter actions for secure
  integrated sensing and communication,'' in \emph{Proc. of IEEE International
  Symposium on Information Theory}, 2024, pp. 2580--2585.

\bibitem{li2025BeamPointing}
S.~Li, M.~Chen, S.~Li, and G.~Caire, ``On secrecy capacity of binary
  beampointing channels with block memory and feedback,'' in \emph{Proc.
  Allerton Conference on Communication, Control, and Computing}, 2025.

\bibitem{alhabob2025Predictive}
A.~A. Al-Habob, O.~A. Dobre, and Y.~Jing, ``Predictive beamforming approach for
  secure integrated sensing and communication with multiple aerial
  eavesdroppers,'' \emph{IEEE Transactions on Communications}, vol.~73, no.~9,
  pp. 7887--7898, 2025.

\bibitem{guo2026secure}
S.~Guo and M.~R. Bloch, ``Secure integrated sensing and communication against
  communication and sensing eavesdropping,'' \emph{arXiv preprint
  arXiv:2601.23216}, 2026.

\bibitem{chen2025distributionPreservingISAC}
Y.~Chen, T.~J. Oechtering, H.~Boche, M.~Skoglund, and Y.~Luo,
  ``Distribution-preserving integrated sensing and communication,'' \emph{IEEE
  Transactions on Information Theory}, vol.~71, no.~10, pp. 7518--7539, 2025.

\bibitem{masouros2026SecureSensing}
K.~Han, K.~Meng, and C.~Masouros, ``Sensing-{Secure ISAC}: Ambiguity function
  engineering for impairing unauthorized sensing,'' \emph{IEEE Transactions on
  Wireless Communications}, vol.~25, pp. 5386--5400, 2026.

\bibitem{ahmadipour2023SecureMessageAndState}
M.~Ahmadipour, M.~Wigger, and S.~Shamai, ``Integrated communication and
  receiver sensing with security constraints on message and state,'' in
  \emph{Proc. of IEEE International Symposium on Information Theory}, 2023, pp.
  2738--2743.

\bibitem{gunlu2023secureBinaryAWGN}
O.~G{\"u}nl{\"u}, M.~Bloch, R.~F. Schaefer, and A.~Yener, ``Secure integrated
  sensing and communication for binary input additive white {Gaussian} noise
  channels,'' in \emph{IEEE International Symposium on Joint Communications \&
  Sensing}, 2023, pp. 1--6.

\bibitem{gunlu2024nonasymptotic}
------, ``Nonasymptotic performance limits of low-latency secure integrated
  sensing and communication systems,'' in \emph{IEEE International Conference
  on Acoustics, Speech and Signal Processing}, 2024, pp. 12\,971--12\,975.

\bibitem{gong2024UntrustedSensing}
Q.~Gong, Y.~Liu, M.~Li, and L.~Ong, ``Secrecy rate-distortion tradeoff for
  integrated sensing and communication with an untrusted sensing node,'' in
  \emph{Proc. of International Conference on Wireless Communications and Signal
  Processing}, 2024, pp. 109--114.

\bibitem{xu2024jammingIRS}
J.~Xu, D.~Li, Z.~Zhu, Z.~Yang, N.~Zhao, and D.~Niyato, ``Anti-jamming design
  for integrated sensing and communication via aerial {IRS},'' \emph{IEEE
  Transactions on Communications}, 2024.

\bibitem{liu2024gameJamming}
Y.~Liu, B.~Zhang, D.~Guo, H.~Wang, G.~Ding, N.~Yang, and J.~Gu, ``A game
  theoretical anti-jamming beamforming approach for integrated sensing and
  communications systems,'' \emph{IEEE Transactions on Vehicular Technology},
  2024.

\bibitem{BlochCovertJCAS}
S.-Y. Wang, M.-C. Chang, and M.~R. Bloch, ``Covert joint communication and
  sensing under variational distance constraint,'' in \emph{Annual Conference
  on Information Sciences and Systems}, 2024, pp. 1--6.

\bibitem{Zhang_2011}
W.~Zhang, S.~Vedantam, and U.~Mitra, ``Joint transmission and state estimation:
  {A} constrained channel coding approach,'' \emph{IEEE Transactions on
  Information Theory}, vol.~57, no.~10, pp. 7084--7095, Oct. 2011.

\bibitem{AcademicsJCASTutorial}
H.~Wymeersch \emph{et~al.}, ``Integration of communication and sensing in {6G}:
  {A} joint industrial and academic perspective,'' in \emph{Proc. of {IEEE}
  Annual International Symposium on Personal, Indoor and Mobile Radio
  Communications}, Helsinki, Finland, Sep. 2021, pp. 1--7.

\bibitem{MassiveMIMOforJCAS}
S.~Buzzi, C.~D'Andrea, and M.~Lops, ``Using {Massive} {MIMO} arrays for joint
  communication and sensing,'' in \emph{Proc. of Asilomar Conference on
  Signals, Systems and Computers}, Pacific Grove, CA, Nov. 2019, pp. 5--9.

\bibitem{MariCaireKramerISACfirst}
M.~Kobayashi, G.~Caire, and G.~Kramer, ``Joint state sensing and communication:
  {Optimal} tradeoff for a memoryless case,'' in \emph{Proc. of IEEE
  International Symposium on Information Theory}, Vail, CO, USA, June 2018, pp.
  111--115.

\bibitem{MariMACJCAS}
M.~Kobayashi, H.~Hamad, G.~Kramer, and G.~Caire, ``Joint state sensing and
  communication over memoryless multiple access channels,'' in \emph{Proc. of
  IEEE International Symposium on Information Theory}, Paris, France, July
  2019, pp. 270--274.

\bibitem{GiuseppeITJSACJournal}
M.~Ahmadipour, M.~Kobayashi, M.~Wigger, and G.~Caire, ``An
  information-theoretic approach to joint sensing and communication,''
  \emph{IEEE Transactions on Information Theory}, vol.~70, no.~2, pp.
  1124--1146, 2024.

\bibitem{AhlswedeCaiWTCwithFeedback}
R.~Ahlswede and N.~Cai, ``Transmission, identification and common randomness
  capacities for wire-tape channels with secure feedback from the decoder,''
  \emph{Electronic Notes in Discrete Mathematics}, vol.~21, pp. 155--159, Aug.
  2005.

\bibitem{AsafCohenWTCwithFeedback}
A.~Cohen and A.~Cohen, ``Wiretap channel with causal state information and
  secure rate-limited feedback,'' \emph{IEEE Transactions on Communications},
  vol.~64, no.~3, pp. 1192--1203, Mar. 2016.

\bibitem{Yingyaosecurefeedback}
Y.~Zhou, N.~Devroye, and O.~Günlü, ``Feedback lunch: Learned feedback codes
  for secure communications,'' in \emph{Proc. {ACM} Workshop on Wireless
  Security and Machine Learning}, Saarbrücken, Germany, July 2026, accepted.

\end{thebibliography}


\begin{IEEEbiography}[{\includegraphics[width=1in,height=1.25in,clip,keepaspectratio]{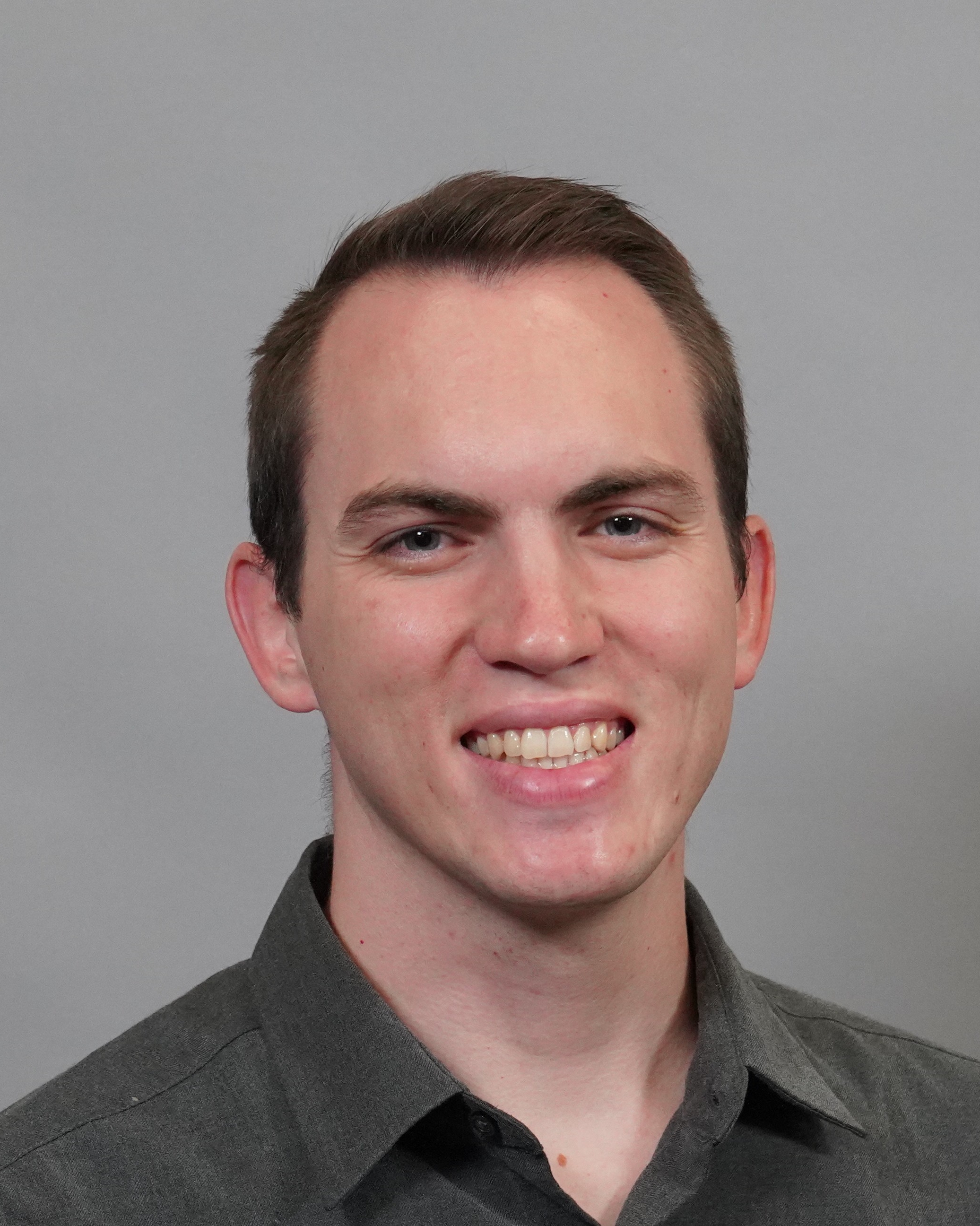}}]{Truman Welling} received a B.S. in Electrical and Computer Engineering with a second major in Applied and Computational Mathematics from Brigham Young University in 2022. He is currently pursuing a Ph.D. degree in Electrical and Computer Engineering as a member of the INSPIRE@OhioState research group at The Ohio State University. His research interests broadly include security in classical and quantum communication systems.
\end{IEEEbiography}

\begin{IEEEbiography}[{\includegraphics[width=1in,height=1.25in,clip,keepaspectratio]{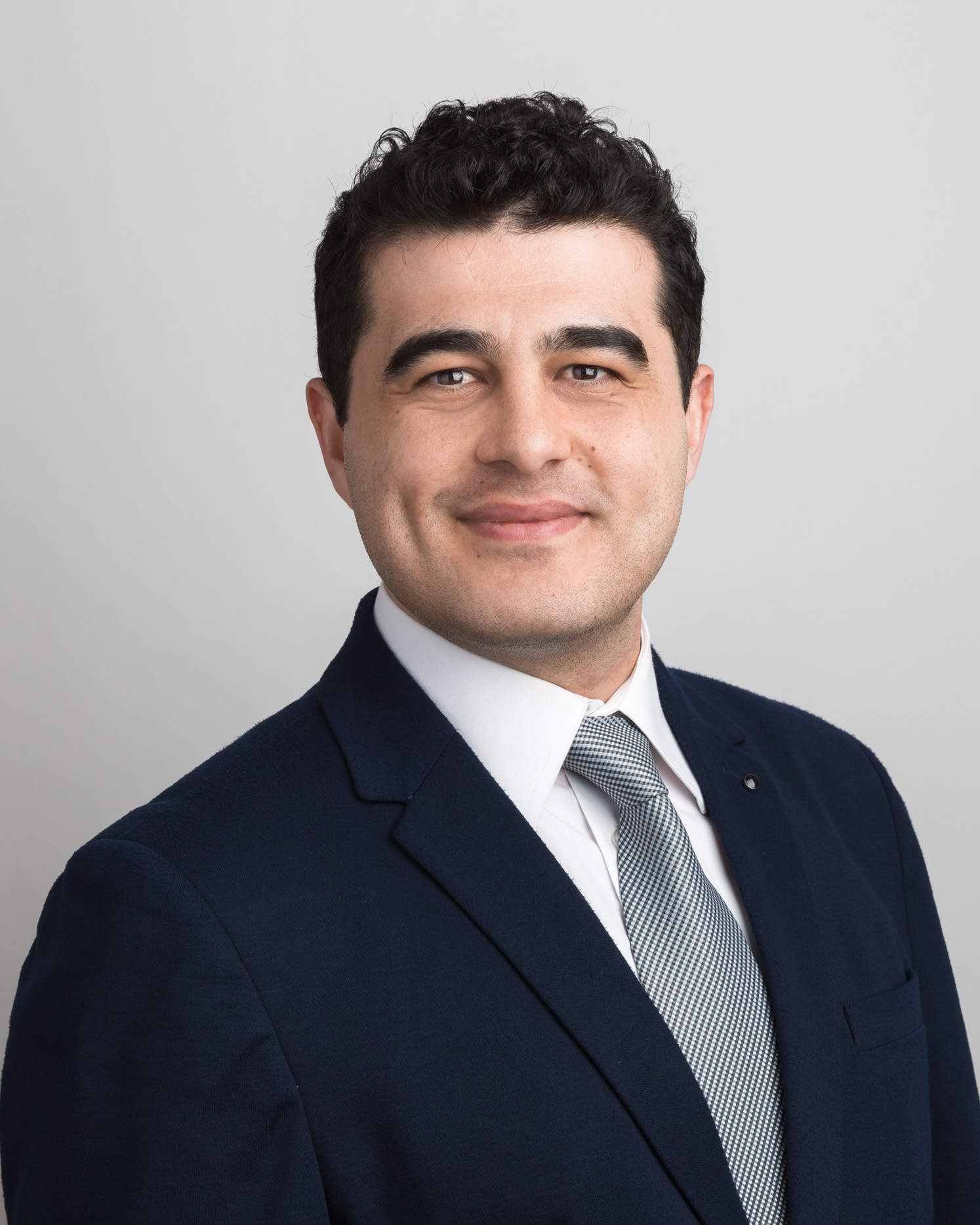}}]{Onur G\"unl\"u} (S'10-M'18-SM'24) received the B.Sc. degree (Highest Distinction) in Electrical and Electronics Engineering from Bilkent University, Turkey in 2011; M.Sc. (Highest Distinction) and Dr.-Ing. (Ph.D. equivalent) degrees in Communications Engineering both from the TU Munich (TUM), Germany in 2013 and 2018, respectively. He was a Working Student in the Communication Systems division of Intel Mobile Communications (IMC), now Apple Inc., in Munich, Germany during November 2012 - March 2013. Onur worked as a Research and Teaching Assistant at TUM between February 2014 - May 2019. As a Visiting Researcher, among more than twenty Research Stays, he was at TU Eindhoven, Netherlands, Georgia Institute of Technology, Atlanta, USA, and TU Dresden, Germany. Following Research Associate and Group Leader positions at TUM, TU Berlin, and the University of Siegen, he joined Linköping University in October 2022 as an ELLIIT Assistant Professor and obtained tenure as an Associate Professor in August 2024. He obtained the Docent (Habilitation) title of Information Theory title in December 2023 and became an IEEE Senior Member in July 2024. Since September 2025, Onur has been a Tenured Full Professor leading the Institute of Communications Engineering at TU Dortmund, Germany and a Guest Professor at Linköping University, Sweden. He has received the 2025 IEEE Information Theory Society - Joy Thomas Tutorial Paper Award, the 2023 ZENITH Research and Career Development Award, 2021 IEEE Transactions on Communications - Exemplary Reviewer Award, and the VDE Information Technology Society (ITG) 2021 Johann-Philipp-Reis Award. His research interests include distributed function computation, information-theoretic privacy and security, coding theory, integrated sensing and communication, and private learning. He serves as an Associate Editor for \textsc{IEEE JSAC}, \textsc{IEEE TCOM}, \textsc{ScienceDirect Journal of Information and Intelligence}, and \textsc{Entropy}. He also serves as a Board Member and Secretary of the IEEE Sweden VT/COM/IT Joint Chapter and as a Working Group Leader for EU COST Action 6G Physical Layer Security (6G-PHYSEC).
\end{IEEEbiography}

\begin{IEEEbiography}[{\includegraphics[width=1in,height=1.25in,clip,keepaspectratio]{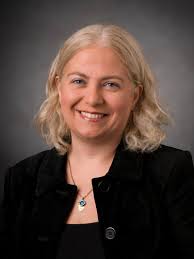}}]{Aylin Yener}
(Fellow, IEEE) received the B.S. degree in electrical and electronics engineering, the second B.S. degree in physics from Bogazici University, Istanbul, Türkiye, and the M.S. and Ph.D. degrees in electrical and computer engineering
from Wireless Information Networks Laboratory, Rutgers University, New Brunswick, NJ, USA. Until 2020, she was a Distinguished Professor of electrical engineering and Dean’s Fellow with Penn State, where she joined as an Assistant Professor in 2002. From 2008 to 2009, she was a Visiting Associate Professor with the Department of Electrical Engineering, Stanford University, Stanford, CA, USA, where she was a Visiting Professor from 2016 to 2017.
She was a Visitor with Telecom Paris Tech, Paris, France, in 2016. Since 2020, she has been the Roy and Lois Chope Chair of engineering with The Ohio State University, Professor with the Department of Electrical and Computer Engineering, Department of Computer Science and Engineering, and the Department of Integrated Systems Engineering. Her expertise is in wireless communications, information theory, and AI, with recent focus on various pillars of 6G, including new advances in physical layer designs, semantic communications, edge learning/computing/AI, system design for confluence of sensing, communications, distributed learning, energy conscious networked systems, and security and privacy. Dr. Yener is a fellow of AAAS and member of the Science Academy of Turkey. She was the recipient of the 2025 IEEE Information Theory Society Joy Thomas Award, 2020
IEEE Communication Theory Technical Achievement Award, 2019 IEEE Communications Society Best Tutorial Paper Award, 2018 IEEE Women in Communications Engineering Outstanding Achievement Award, 2014 IEEE Marconi Paper Award, and several other research and technical awards. Yener is the Chair-elect for the IEEE Technical Activities Board, and Vice President Elect for Technical Activities (which contains all IEEE Society and Councils) of the IEEE. From 2024 to 2025, she was on the IEEE Board of Directors as the Director of Division IX. In 2020, she was the President of IEEE Information Theory Society. She is the Editor-in-Chief of IEEE TRANSACTIONS ON GREEN COMMUNICATIONS AND NETWORKING, and Senior and Guest Editor of numerous IEEE journals. She has been the Co-Founder of the IEEE North American School of Information Theory which runs annually, since 2008.
\end{IEEEbiography}

\end{document}